# A Probe into Causes of Non-citation Based on Survey Data

Zewen Hu [a]   Yishan Wu [b*]

[a] School of Economics and Management, Nanjing University of Information Science & Technology, Nanjing, Jiangsu 210044, China

[b] Chinese Academy of Science and Technology for Development, Beijing 100038, China

**Abstract**   Empirical analysis results about the possible causes leading to non-citation may help increase the potential of researchers' work to be cited and editorial staffs of journals to identify contributions with potential high quality. After complete literature retrieval from the Web of Science, Google Scholar, and Scopus databases, we find that very few studies focus on the empirical analysis of causes of non-citation.

In this study, we conduct a survey on the possible causes leading to citation or non-citation based on a questionnaire. We then perform a statistical analysis to identify the major causes leading to non-citation in combination with the analysis on the data collected through the survey. Most respondents to our questionnaire identified eight major causes ( 'research hotspots and novel topics of content', 'longer intervals after publication', 'research topics similar to my work', 'high quality of content', 'reasonable self-citation', 'highlighted title', 'prestigious authors', and 'academic tastes and interests similar to mine' ) that facilitate easy citation of one's papers.

They also pointed out that the vast difference between their current and former research directions as the primary reason for their previously uncited papers. They feel that text that includes notes, comments, and letters to editors are rarely cited, and the same is true for too short or too lengthy papers. In comparison, it is easier for reviews, articles, or papers of intermediate length to be cited.

---

[*] Address for correspondence:
Yishan Wu
Permanent address: Chinese Academy of Science and Technology for Development, No.8 Yuyuantan South Road, Haidian District, Beijing 100038,China.
 Tel.: +86 010 13501127855.
  E-mail address: wuyishan1958@163.com

# 1. Introduction

'Citation of papers' represents a positive evaluation of the quality of cited papers and the related publication entities including scientific journals, research groups, individual scientists, research institutes, universities, and countries (Nisonger 2004; Castellano and Radicchi 2009), while 'non-citation of papers' sometimes represents a negative evaluation, In sum, the value and quality of papers and the related publication entities are important causes of whether a paper can get cited easily or not. Of course, there are also some papers that even when not cited in a long timeframe, are still of high potential value, making it possible for them to be cited frequently in the future timeframe. Such papers are called 'sleeping beauties' by some authors (Van Raan, 2004; Burrell 2005). Many years ago, some scholars theoretically explained the reasons of the 'non-citation phenomenon'. Garfield (1973) considered that uncited papers may be due to the fact that they are mediocre, of low quality, published in low-impact source journals, unintelligible, irrelevant, valuable but undiscovered or forgotten, and so on (Garfield, 1973). Garfield (1991) further listed a series of possible reasons leading to non-citation of articles, such as language, type of publication, being 'premature', delayed recognition, bibliographic plagiarism, or other variations of misconduct. However, these descriptions about the reasons of non-citation were seldom supported by empirical studies.

Fortunately, many years later, a series of efficient empirical studies that focus on the influencing factors of non-citation were executed by van Leeuwen and Moed (2005), Egghe (2008, 2010, & 2013), Hsu and Huang (2012), and Burrell (2013), who have revealed the decreasing S-shape function relationship between impact factor of journals and the non-citation factor. Furthermore, a positive correlation between the h-index and the number of uncited papers have also been verified on a sample of 75 top researchers from the fields of mathematics (fields medalists), physics, chemistry, and physiology or medicine (Nobel laureates). A paper issued by Li J and Fred Y. Ye in 2014 made an effort to probe different citation patterns of high-quality and high-impact publications through an empirical analysis, which indicate that the quality and impact of publications do influence citation of papers (Li J. & Fred Y. Ye, 2014). This judgment seems contradictory to the opinions of other scholars. For example, Glänzel et al (2006) said: 'The fact that a document is less frequently cited or even (still) uncited during several years after publication provides information about its reception by colleagues but does not reveal anything about its quality or the standing of its author(s) in the community'. Hu and Wu (2014) further conducted an empirical pilot study that found the length of a paper has a great influence on whether it would be cited or not (Hu & Wu, 2014).

Understanding the reasons or influencing factors of non-citation may help such entities as scientific journals, researchers, institutes, universities, and countries improve the chances of getting their papers cited and lower their percentage of uncited papers to raise their performance in impact assessments and research quality. However, the current literature focuses more on the 'citation phenomenon', while the 'non-citation phenomenon' hardly gets serious attention of scholars. Through a complete literature retrieval from the Web of Science, Google Scholar, and Scopus databases, we find that until now, most scholars tend to make some studies about how to model mutual relationship between impact factor and the non-citation factor, and there has not been any thorough empirical study that analyses the possible reasons leading to non-citation of papers by way of questionnaire surveys.

**2. Methodology**

For thoroughly verifying possible reasons leading to non-citation, we firstly design a questionnaire by reviewing all related literature and consulting some related experts, researchers, and authors. Then we statistically analyse the data obtained through the questionnaire, and present and discuss the most important reasons as recognized by most researchers or readers of papers, as well as the unrecognized reasons.

**2.1. Content of the questionnaire**

In the questionnaire, we design a series of questions that can reflect respondents' views and attitudes to various reasons leading to non-citation. Respondents mainly comprise researchers or readers of papers from different domains and with different education backgrounds. Table 1 shows all the questions related to reasons of citation or non-citation. To some extent, reasons of citation are also in turn reasons of non-citation.

Table 1 Questions related to reasons of citation or non-citation

| NO | Questions |
|---|---|
| Q0 | Do you know whether your papers are cited? |
| Q1 | Do you know whether you have any uncited papers? |
| Q1.1 | If you have uncited papers, what do you think are the main reasons leading to their non-citation? |
| Q1.2 | If all your papers are cited, what do you think are the reasons that lead to their citation? |
| Q2 | If you have some papers that have not been cited until now, why don't you change them into cited papers through self-citation, thus making a breakthrough from non-citation to citation? |
| Q3 | Which type of papers do you think are not being easily cited? |
| Q4 | If the number of words in a paper influences its potential of being cited or not, ideally, how many words for a paper do you think can lead to its easier citation? |
| Q5 | Have you encountered situations, where you see a paper of good quality is not cited or less cited, while a paper with mediocre or low quality is cited more often? |
| Q6 | Do you agree to the view that a paper that is not cited in a citation time window of ten years has no value? |
| Q7 | What do you think are the reasons leading to citations of others' papers? |

In Table 1, we list ten questions that explore the reasons of citations or non-citations. In these questions, Q1.1 and Q1.2 are two sub-questions of Q1.

**2.2. Process and results of the questionnaire survey**

An online questionnaire containing ten questions (Table 1) was designed and uploaded on related websites, blogs, forums, and so on, on 1 November 2014 and kept open for six months. It attracted the attention of 277 respondents, of which 274 (98.92%) respondents completed the questionnaire. However, by checking respondents' IP addresses, names, and contact information, we found that some questionnaires were repeatedly answered. After deleting the repeated ones, 240 effective questionnaires were retained as a research sample, reducing the effective rate of questionnaires to 87.59%.

By a statistical analysis of responses to question that is called Q0 in Table 1, we found that 198（82.5%）participants know the citation situation of their papers, while 42（17.5%）respondents are not aware of the same. Further, we statistically analyse 198 questionnaires filled by 198 participants knowing their citation situation.

Major domain distribution, degree and title distribution, and research output distribution of 198 participants are shown in Figure 1, Table 2, and Table 3, respectively.

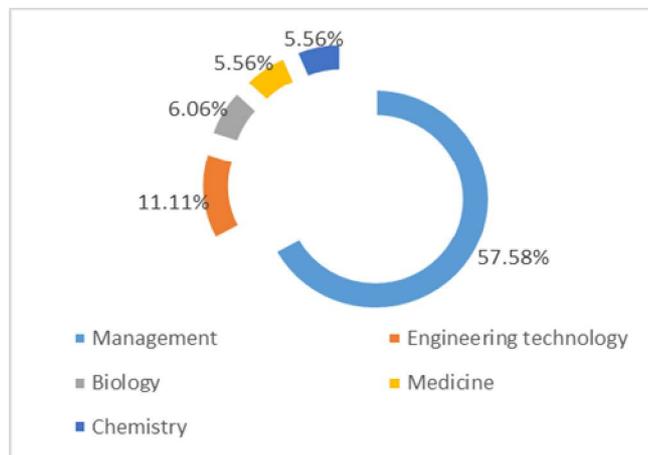

Figure 1 Domain distribution of 198 participants

Figure 1 shows the domain distribution of 198 participants from various education backgrounds. 114 (57.58%) participants with management background and 22 (11.11%) participants from engineering technology expressed their views or attitudes to various reasons of non-citation, while there are only 12 (6.06%), 11(5.56%), and 11(5.56%) participants from the domains of biology, medicine and chemistry.

Table 2 Degree and title distribution of 198 participants

| Degree | Percentage | Title | Percentage |
| --- | --- | --- | --- |
| Bachelor | 6.06% | Assistant Lecturer & Lecturer | 11.62% |
| Master | 42.93% | Senior Lecturer | 35.86% |
| Doctor | 42.93% | Associate Professor & Professor | 29.29% |
| Postdoctoral | 7.07% | No Title | 23.23% |

From Table 2, we can see that among the 198 participants who answered the questionnaire, most participants have masters or doctoral degree, as well as the titles of senior lecturer or professor. A total of 170 participants hold masters or doctoral degrees (85.86%). Furthermore, 70 (35.86%) hold the title of senior lecturer and 58 (29.29%) hold the title of associate professor and professor.

Table 3 Research output distribution of 198 participants

| Research output | Number of participants | Percentage |
|---|---|---|
| ≤5 papers | 66 | 33.50% |
| 6-10 papers | 38 | 19.29% |
| 11-15 papers | 41 | 20.81% |
| ≥16 papers | 52 | 26.40% |

As shown in Table 3, among the 198 participants who answered the questionnaire, 40.1% of the participants have research outputs of six to fifteen papers, 26.40% of the participants have more research outputs of more than or equal to sixteen papers. While the other 33.50% of the participants have research outputs of less than or equal to five papers.

3. Results

Based on the questionnaire data answered by 198 participants knowing their citation situation from different domains and with different degrees, titles, and research outputs, we make a statistical analysis of questions 1-7 in Table 1.

3.1. Analysis results of Q1 and its two sub-questions, Q1.1 and Q1.2

Through a statistical analysis of Q1 and its two sub-questions, Q1.1 and Q1.2, we find that among the 198 participants, 156（78.79%）participants indicate that they have papers that have not been cited until 30 April, 2015 and express their views and attitudes to the reasons of non-citation, as shown in Table 4, while 41（20.7%）participants indicate that they do not have uncited papers, and express their views and attitudes to the reasons for the same, as shown in Table 5.

**Table 4 Views and attitudes of 156 participants to reasons leading to non-citation of papers**

| Views and attitudes | Number of participants with approval | Percentage |
|---|---|---|
| Shorter online time of papers (less than two years) | 86 | 54.78% |
| Lower quality of content | 77 | 49.04% |
| Older or colder topics | 72 | 45.86% |
| Source journals with lower impact or quality | 63 | 40.13% |
| Authors of papers with lower popularity in the science world | 60 | 38.22% |
| No active propagation and recommendation of papers | 41 | 26.11% |
| Some papers are inevitably ignored among a large number of papers | 20 | 12.74% |
| Views of papers are too recent and advanced to be understood | 14 | 8.92% |
| English papers by Chinese authors always attract lesser attention | 11 | 7.01% |
| Other reasons | 9 | 5.73% |

In Table 4, it is to be noted that among 156 participants with non-cited papers, shorter publication time, lower quality of content, and older or colder topics of papers are considered by 86 (54.78%), 77 (49.04%), and 72 (45.86%) participants respectively, as the major reasons leading to non-citation. While only 14 (8.92%) and 11 (7.01%) participants consider the following reasons of non-citation of their papers: views of papers being too recent and advanced to be cited, and international scholars always paying lesser attention to English papers by Chinese authors.

**Table 5 Views and attitudes of 42 participants to reasons for all their papers being cited**

| Views and attitudes | Number of participants with approval | Percentage |
|---|---|---|
| Hot topic | 29 | 70.73% |
| High quality of content | 16 | 39.02% |
| Reasonable self-citation | 16 | 39.02% |
| Often recommend my papers to peers | 9 | 21.95% |
| Prestigious co-authors | 5 | 12.20% |
| Others | 4 | 9.76% |
| Being a tutor, my students are willing to cite my papers | 1 | 2.44% |

From Table 5, we can see that the 42 participants with all-cited papers shared the following various reasons about why all their papers were cited: 29 (70.73%) participants think it is their selection of hot topics, 16 (39.02%) participants think it is the high quality of content and reasonable self-citation; however, only 5 (12.20%) and 1(2.44%) participants consider their famous co-authors and their students who are willing to cite their papers as the reasons.

**3.2. Analysis results of question Q2**

Among 198 participants, 156 participants with uncited papers until now express their views and attitudes to reasons they do not make a breakthrough from non-citation to citation by self-citation, as shown in Table 6

**Table 6 Views and attitudes of 156 participants to reasons they do not make a breakthrough from non-citation to citation by self-citation**

| Views and attitudes | Number of participants with approval | Percentage |
|---|---|---|
| I can't cite my previous uncited papers as there is a great difference between my current research direction and | 93 | 59.62% |
| previous research direction | | |
| Self-citation may be seen as boasting about one's papers | 56 | 35.90% |
| I don't want to cite my uncited papers of low quality | 48 | 30.77% |
| I never consider the option of citing my own papers | 44 | 28.21% |
| I have not been engaged in research | 21 | 13.46% |
| Others | 22 | 14.10% |

From Table 6, we can see that among the reasons of not making a breakthrough from non-citation to citation by self-citation, the major reason is that they can't cite previous uncited papers due to the great difference between their current research direction and previous research direction; 59.62% participants agree with this view. Whereas 35.90% and 30.77% of participants worry about being criticized for boasting or consider their quality of papers too low to be self-cited.

### 3.3. Analysis results of question Q3

Regarding the question of whether types of papers influence their future chances of being cited, Garfield (1991) considered types of publications as a factor that influences their citation chance. However, Garfield's view has not been verified by empirical data. Therefore, an empirical survey to question Q3 was carried out, and the analysis results are shown in Table 7.

**Table 7 Views and attitudes of 198 participants to types of paper that are not easily cited**

| Views and attitudes | Number of participants with approval | Percentage |
|---|---|---|
| Types of papers such as notes, comments, and letters | 81 | 40.91% |
| Empirical analysis papers | 76 | 38.38% |
| Theory research papers | 47 | 23.74% |
| Reviews | 31 | 15.66% |

Table 7 shows some meaningful views. 40.91% and 38.38% of the participants feel that certain types of papers such as notes, comments and letters and papers on empirical analysis hold lesser chances of being cited. However, only 23.74% and

15.66% of the participants feel that papers on theory research and reviews are not easily cited.

**3.4. Analysis results of question Q4**

The question about the optimum number of words in a paper that can improve the number of times it is cited is an interesting one. In a previous study (Hu & Wu, 2014) we concluded that the length of a paper has a great influence on whether it will be cited by analysing the citations of papers with different lengths in six sample journals. Generally, shorter papers with 1-4 pages have a smaller chance of being cited, longer papers with 5 and more pages have a greater chance of being cited. In this study, we survey the views and attitudes of scholars, researchers, and research students concerning the optimum number of words in a paper that will improve its chances of being cited, and we contrast our results with that from previous analyses. The views and attitudes to this question from 198 participants are as shown in Table 8. The number of English words is determined by the following equation: 1 English word = 2 Chinese words.

**Table 8 Survey on the optimum number of words that improves a paper's chances of being cited**

| Number of words | Number of participants | Percentage |
| --- | --- | --- |
| 2000-4000 words | 126 | 63.64% |
| 4000-7500 words | 67 | 33.84% |
| 1000-2000 words | 38 | 19.19% |
| More than 7500 words | 28 | 14.14% |
| 500-1000 words | 8 | 4.04% |

Some meaningful inferences can be made from the information in Table 8: a majority of the participants (63.64%) agree that papers with 2000-4000 words have a greater chance of being cited by other scholars. Next, 33.84% of the participants agree that papers with 4000-7500 words have a greater chance of being cited. Finally, only 4.04% of the participants think that papers with 500-1000 words have a great chance of being cited by others. A conclusion similar to our previous statistical analysis result on the length of papers and their chances for citation can be obtained: that shorter papers have a smaller chance of being cited and the longer papers have a greater chance.

### 3.5. Analysis results of question Q5

General views on citation are that high-quality papers are cited more often, while low-quality papers may have a very small chance of being cited. To verify such views, we carry out a survey on the topic. Survey data and analysis results show that 129 (65.15%) of the 198 participants have encountered situation in which good-quality papers as their own understanding are not cited or are cited fewer times, while a low-quality paper are cited more times. The other 69 (34.85%) participants have not encountered such a situation.

### 3.6. Analysis results of question Q6

According to the theory of literature obsolescence, the older the literature is, the lower its value becomes, subsequently lowering the chance of citation. Does that mean the literature that is not cited during 10 years following their publication has no value? We posed this question in a survey to 198 participants. Survey results show that 148 (74.75%) participants do not identify with the view that the literature uncited during 10 years following their publication has no value. The other 50 (25.25%) participants supported this view.

### 3.7. Analysis results of question Q7

Why are you willing to cite this paper and not another paper? You must have your reasons. Here we carry out a survey to understand these reasons in citing other researchers' papers. The views and attitudes to this question of 198 participants are shown in Table 9.

**Table 9 Views and attitudes on reasons for citing others' papers**

| Views and attitudes | Number of participants | Percentage |
|---|---|---|
| Higher quality of content | 161 | 81.31% |
| Similarity of topics | 158 | 79.80% |
| Highlighted title of paper | 119 | 60.10% |
| Topics of cited paper are novel | 118 | 59.60% |
| Similarly academic tastes and interests | 112 | 56.57% |
| Popularity of cited authors is higher | 91 | 45.96% |
| Need to cite some papers from journals in which I wish to publish my paper | 64 | 32.32% |
| Higher impact factor for source journals of cited papers | 62 | 31.31% |
| Maintain good relationships with researchers in the same academic domain | 51 | 25.76% |
| Cited papers have some gaps that I want to criticize | 46 | 23.23% |
| Summary of cited papers has higher correlation with my paper | 45 | 22.73% |
| I cite papers that are copied from another paper's reference list without reading them | 22 | 11.11% |
| Cited authors are reviewers of journals in which I wish to publish my paper. | 19 | 9.60% |
| Other | 5 | 2.53% |

As shown in Table 9, high quality of content, similar research topics, and highlighted title are the main reasons for citing a paper. More than 111 (56%) of 198 participants are willing to cite a paper on account of its higher quality of content (81.31%), similar research topic (79.80%), and highlighted title (60.10%), novel topics of paper (59.60%) and similarly academic tastes and interests (56.57%).

Some participants would cite papers for the following reasons: 91 (45.96%) chose because the author is popularly cited , 64 (32.32%) thought it important to ingratiate oneself with the editorial department of journals by citing papers from their journals, 62 (31.31%) wished to cite papers from high-impact-factor journals, 51 (25.76%) would cite authors to maintain good relationships with those in their academic domain, 46 (23.23%) wished to cite papers to highlight the gaps in research.

Few participants have the least effort principle and utilitarian purposes as reasons for citing papers. For example, only 19 (11.11%) participants act on the least effort principle and cite some related paper that they have never read from another paper's reference list as reference. Nineteen (9.60%) participants would cite papers because the cited authors are reviewers of journals in which they wish to publish their paper.

## 4. Conclusion

Scholars have theoretically examined the reasons for non-citation. Empirical studies that focus on the influencing factors of non-citation have been executed by van Leeuwen and Moed (2005), Egghe (2008, 2010, & 2013), Hsu and Huang (2012), and Burrell (2013), and these mainly verify the decreasing function relationship between impact factor, h-index, and the uncitedness factor. However，there have been no empirical studies that discuss the most important reasons or influencing factors as recognized by many researchers or readers of papers, as well as the unrecognized reasons or influencing factors. We conducted such a pilot empirical analysis on possible reasons of non-citation based on data collected via a questionnaire survey. The following significant conclusions are drawn.

- Our online questionnaire received a good response. A total of 277 questionnaires were distributed, and 274 questionnaires were completed, with the response rate at 98.92%. After removing duplicating and invalid questionnaires, 240 effective questionnaires were obtained, setting the response rate at 87.59%.
- Of the 240 questionnaires, we selected 198 participants who are aware of the citation situation of their papers as our sample. Of the 198 participants, 57.58% of the participants are from the field of management, and 11.11% of the participants are from engineering technology. Furthermore, 85.86% of the participants have masters degrees or doctoral degrees, and 35.86% of the participants are senior lecturers. The percentage of participants who have published 6 or more papers is 66.50%.
- Of the 156 participants who have papers that have not been cited as of April 30, 2015, more than 45.86% consider ''shorter publication time', 'lower quality of content', 'older or colder topics of papers' as major reasons for their papers not being cited. Less than 9% of the participants attribute the non-citation of their papers to avant-garde topics and less attention from international scholars. Furthermore, the major reasons that many of the participants (more than 30%) do not make the breakthrough from non-citation to citation by self-citation include large differences between their current and former research, worry about being seen as narcissistic, or low quality of their previous papers.

- Among 42 participants who have papers that have not been cited, above 39.02% think that hot topics, high quality, and reasonable self-citation improve a paper's chances of being cited.

- Among 198 participants, 40.91% and 38.38% of them think that there is a very small chance of citation for the following paper types: empirical analysis, notes, comments, and letters. On the contrary, theory research and reviews have a higher chance of being cited.

- Of the 198 participants, 63.64% agree that papers with 2000-4000 words have a greater chance of being cited, while 33.84% think that there is a greater chance of citation for those papers with 4000-8000 words. Less than 20% of the participants think that there is a great chance of citation for papers with 500-2000 words or more than 7500 words. Furthermore, more than 129 (65.15%) of the 198 participants have encountered a situation in which a good-quality paper is not cited or is cited only a few times, and they do not identify with the view that literature uncited after 10 years of its publication has no value.

- More than 119 (56%) of the 198 participants would cite a paper on account of its higher quality of content (81.31%), similar research topic (79.80%), highlighted title (60.10%), novel topic of the paper (59.60%), and similar academic tastes and interests (56.57%). Less than 12% of the 198 participants would have the least effort principle (citing papers they have not read by copying from another paper's reference list) and utilitarian purposes (citing authors who are reviewers of journals in which they wish to publish their paper) as reasons for citing a paper.

The questionnaire survey is an effective method of analysing the reasons for non-citation. However, compared to such complex methods for analysing influencing factors as structural equation model analysis and panel data model analysis, a statistical analysis based on a questionnaire survey can't model relationships among non-citation and its causes or influencing factors. Therefore, in the future, we will design a structural equation model for such relationships by executing a more comprehensive questionnaire covering all kinds of influencing factors of non-citation, such as 'contents and topics of papers', 'academic status of journal', and 'bibliometric characteristics of papers'.

Acknowledgments
This study is supported by the National Natural Science Foundation of China (Grant No. 71373252).